# Females' Enrollment and Completion in Science, Technology, Engineering, and Mathematics Massive Open Online Courses

Suhang Jiang, University of California, Irvine, suhangj@uci.edu
Katerina Schenke, University of California, Los Angeles, kschenke@ucla.edu
Jacquelynne Sue Eccles, University of California, Irvine, jseccles@uci.edu
Di Xu, University of California, Irvine, dix3@uci.edu
Mark Warschauer, University of California, Irvine, markw@uci.edu

## Abstract

Massive Open Online Courses (MOOCs) have the potential to democratize education by providing learners with access to rich sources of information. However, evidence supporting this democratization across countries is limited. We explored the question of democratization by investigating whether females from different countries were more likely to enroll in and complete STEM MOOCs compared with males. We found that whereas females were less likely to enroll in STEM MOOCs, they were equally likely to complete them. We found smaller gender gaps in STEM MOOC enrollment in less economically developed countries. Further, females were more likely than males to complete STEM MOOCs in countries identified as having a high potential to become the largest economies in the 21st century.

## Introduction

Massive Open Online Courses (MOOCs) have attracted tens of millions of learners around the world. Theoretically, anyone with an Internet connection is able to freely access these online courses, which are often provided by professors from elite universities. Similar to previous technological advancements in broadcast media such as radio and television, MOOCs were expected to transform education by providing learning opportunities for those who otherwise would not have access to them (Yuan & Powell, 2013). The growing MOOC movement stems from the belief that knowledge should be freely shared and people have the right to learn regardless of their social and economic backgrounds (Yuan & Powell, 2013). MOOC proponents argue that MOOCs can democratize higher education and provide learning opportunities not only for traditionally underserved populations but also for college-educated populations, who may benefit from the extra course-work made available to improve their employment opportunities (Koller, 2013).

However, the high expectation that MOOCs will promote educational equity has been dampened by studies describing the demographics of individuals who enroll in and complete MOOCs (Christensen et al., 2013; Hansen & Reich, 2015; Ho et al., 2015). Statistics show that the majority of MOOC learners are young, well-educated males from developed countries (Christensen et al., 2013). In the U.S. for example, people of higher socioeconomic status (SES) are much more likely to enroll in MOOCs than people of lower SES (Hansen & Reich, 2015). Gender disparity is also prevalent in MOOCs, especially in STEM subjects. On average, only 1 in 5 learners in a STEM MOOC is female (Ho et al., 2015). Based on these demographics, critics argue that MOOCs are failing to reach disadvantaged individuals, such as those without access to higher education in developing countries (e.g., Emanuel, 2013). This critique implicitly assumes that those who already have a college degree in developing countries should not be considered as disadvantaged. However, compared to their peers from developed countries, those who already have a college degree in developing countries are still at a disadvantage in terms of accessing high quality education from elite universities as well as high-quality jobs that result from this elite education.

In addition to critiques about MOOCs reaching disadvantaged individuals, criticism has been voiced about whether MOOCs increase the participation of females in STEM fields (Ho et al., 2015). We are particularly interested in females' enrollment and performance in STEM MOOCs, as females have been traditionally underrepresented in STEM fields. For example, females constitute 29% of science and engineering occupations in the United States (Beede et al., 2011), 12.8% in the UK (Arnett, 2015), 16% in Australia (Office of the Chief Scientist, 2016), and 13.8% in Japan (Homma, Motohashi, & Ohtsubo, 2013). Increasing female's participation in STEM fields is crucial for strengthening the STEM workforce and a country's global competitiveness (Beede et al., 2011). Though females are in general underrepresented in STEM MOOC participation, it is unclear whether the gender disparity differs across countries. No studies have explored how the country-level disadvantage combined with individual-level disadvantage is associated with an individual's enrollment and completion in MOOCs.

This study will explore the association between macro societal factors and females' enrollment in and completion of STEM MOOCs. Our analytical framework is guided by the Eccles' Expectancy-Value Model of Achievement-Related Choices (Eccles, 1994; Wigfield & Eccles, 2000). This model suggests that social context and cultural forces contribute to gendered educational choices (Eccles, 1994; Wigfield & Eccles, 2000). In addition, we consider economic advancement in influencing females' choices to enroll in and complete STEM MOOCs because, on average, economic advancement is associated with less gender segregation and more gender equality in education (Baker & Jones, 1993). In the context of the enrollment and completion of STEM MOOCs, it is not obvious a priori what the direction of the relationship will be. The percentage of STEM students who are female is higher for countries higher in gender equality (van Langen & Dekkers, 2005). For instance, in Iceland, which prioritizes gender equality, a relatively high number of females are in science and engineering education (Hanson, Schaub, & Baker, 1996). In addition, previous studies show that there are smaller gender differences in math performance in more gender-equal cultures (Guiso, Monte, Sapienza, & Zingales, 2008; Hyde & Mertz, 2009). If this pattern were the norm, we would expect a smaller gender gap in STEM MOOC enrollment and completion in more gender-equal and economically developed countries than from other countries.

However, some studies find greater gender segregation in education fields in more economically developed countries. For instance, Finland is found to have the highest level of gender segregation in fields of study among 44 countries (Charles & Bradley, 2009). The cultural beliefs that males and females are fundamentally and innately different and the opportunities to express a gendered identity may account for the pronounced gender segregation in socially and economically developed countries (Charles & Bradley, 2009). Females from developed countries may feel that it is legitimate to express their aversion to math or STEM-related courses, which reinforces their inclination not to choose STEM fields. If this were the case, we would expect more gender segregation in STEM MOOCs in more developed countries than in other countries.

Even in developing countries, there is a positive association between GDP and gender segregation in fields of study (Charles & Bradley, 2009). Based on this, we may expect that for developing countries, higher economic development will be associated with higher gender segregation, i.e., females are less likely to enroll in and complete STEM MOOCs than males.

The lack of access to high quality STEM courses is one of the factors that hinders students' enrollment in traditional STEM fields (van Langen & Dekkers, 2005), and this may be especially true for females from developing countries. In addition, females in fast-changing economies (e.g., Mexico) may enter STEM fields (e.g., Computer Science) to maximize their earnings (Khazan, 2014). For instance, 77% of female respondents from developing countries stated that they feel encouraged to work in STEM fields while only 46% of female respondents from developed countries did (Penn, 2015).

The general conservative social norms and cultural expectations in many developing countries may decrease the possibility that females will choose STEM courses. Nevertheless, the free and easy access to the online courses provided by elite universities may spark females' interest in STEM fields and facilitate course enrollment, despite social expectations and cultural scripts. If true, females from less developed countries might be more interested in taking advantage of STEM MOOCs.

This study explores the association between nation-wide gender equality, economic development, and females' enrollment in and completion of STEM MOOCs. We specifically look at these two outcomes separately because MOOCs are notorious for having very low completion rates (Ho et al., 2015). Additionally, different factors may be associated with whether an individual decides to enroll in a STEM MOOC and whether an individual actually completes it.

**Materials and methods**
To address our research questions, we used the HarvardXMITx Person-Course de-identified dataset from the 2012-2013 academic year (Fall 2012, Spring 2013, Summer 2013), which included 16 HarvardX and MITx courses on the edX platform constituting the most comprehensive public dataset on MOOCs. Thirteen MOOCs were labeled as STEM MOOCs and three MOOCs were labeled as non-STEM (see table S1 in the supplementary materials for the complete list including platform, course title, percent females, and whether STEM or non-STEM). In the present study, STEM MOOCs were defined as follows: Biology, Computer Science, Engineering and Mechanics, Physics and Chemistry, and Public Health. Learners in these online courses were from across the world. Demographic information for the sample can be found in the supplementary materials. The dataset included user self-reported variables such as gender, age, highest level of education, country, and information about the courses that they enrolled in and whether they have completed those courses. There were 641,138 person-course observations in the original dataset. We removed observations with missing gender information (86,823) and those that did not have specific country names (175,370), such as "other Europe". We then aggregated the person-course dataset (378,945); the dependent variable was set to 1 if the participant took at least one STEM MOOC and 0 if the participant did not take a single STEM MOOC. Using this method, we obtained 293,144 unique learner observations from 25 countries. Table S2 in the supplementary materials contains information on the gender composition of STEM MOOC enrollment for each country in the final dataset.

We used the Gender Gap Index (GGI) created by World Economic Forum to measure a country's gender equality level (World Economic Forum, 2012). GGI reflects the gap between males and females in access to resources and opportunities of health, education attainment, economic participation and political empowerment and was used as a key predictor variable in our models. The GGI was composed of the country's health index, education attainment index, economic participation index, and political empowerment index. The health index refers to the sex ratio at birth and the gap between females' and males' healthy life expectancy. The education attainment index reflects the ratios of females to males in primary-, secondary-, and tertiary-level education. Economic participation index reflects the gap between females' and males' labor force participation rate, wage equality, and the ratio of females to males among professional workers and senior officials. The political empowerment index reflects the gap between females and males at the highest-level of political decision making (World Economic Forum, 2012). GGI ranges from 0 (inequality) to 1 (equality) and indicates the reduced gap between males and females for an individual country. The higher the GGI refers to the more gender-egalitarian environment. The GGI for the 25 countries in the dataset ranges from 0.55 (Pakistan) to 0.78 (Philippines). For this study, we center GGI on the lowest value of GGI, i.e. 0.55. As GGI does not reflect a country's development level, we included the gross domestic product (GDP) per capita (2012) to measure a country's economic development level [24]. The GDP per capita for the countries in the dataset ranged from $859 (Bangladesh) to $67,512 (Australia). We also included controls for the learner's age (mean= 28.65, SD=0.02, range 10 to 82 years old; individuals who reported being younger than 10 years old were labeled as missing) and education-level measured on a 5-point scale where 1 was less than Secondary, 2 was Secondary, 3 was Bachelor's degree, 4 was Master's degree, and 5 was Doctorate degree. The mean education level for the analysis sample was 2.93 with a standard deviation of 0.02 and a range of 1 to 5. Because descriptive analyses of STEM MOOC completion rates across countries showed that several top countries with the highest female representation in STEM MOOC completion were Next Eleven countries (which have been identified as having a high potential to become the largest economies in the 21st century (O'Neil, Wilson, Purushothaman, & Stupnytska, 2005)), we also explored females' completion patterns within the Next Eleven countries (Bangladesh, Egypt, Indonesia, Nigeria, Pakistan, the Philippines).

To answer our research questions on the association between nation-wide gender equality, economic development, and females' enrollment in and completion of STEM

MOOCs, we conducted a series of logistic hierarchical linear models to account for nesting of individual within country. We tested for the inclusion of random slopes and random intercepts all of our models. Using the likelihood ratio test, we found that random slope models performed significantly better than models where only the intercept was allowed to vary randomly random; therefore we report results from models were slopes were able to vary randomly. To examine the degree to which learners from different countries differ in their propensity to choose and complete STEM MOOCs, we calculated the intraclass correlation coefficient (ICC) to determine if there was sufficient country-level variance to model (Snijders & Bosker, 2011). The ICC is .21 for enrollment and .13 for completion, indicating that about 21% and 13% of the variation in STEM MOOC enrollment and completion respectively can be attributed to differences in learners' country of origin. We first ran multilevel logistic regression models for STEM MOOC enrollment and STEM MOOC completion separately. Finally, we present results for STEM MOOC completion for both individuals from Next Eleven countries and from non Next Eleven countries to further understand the association between gender, nation-wide gender equality and STEM MOOC completion in Next Eleven countries.

**Results**

Figure 1 displays the gender composition in STEM MOOC enrollment across countries. Females comprised only 23.89% of STEM MOOC learners in the dataset. By country, the percentage of STEM MOOC learners who were female ranged from 5.25% in Bangladesh to 38.32% in Philippines. It is worth noting that three of the top four countries of highest female representations were developing countries, i.e., Philippines, Indonesia and Colombia. Figure 2 shows the percentage of MOOC enrollees in each country taking at least one STEM MOOC by gender. On average, 70.74% of females in the dataset chose to enroll in at least one STEM MOOCs. The percentage of female MOOC enrollees taking at least one STEM course, ranged from 17.14% in Japan to 96.99% in Portugal. Figure 2 also shows that female and male enrollees took STEM courses at nearly the same rate in many countries including Portugal, Egypt and Nigeria. For example, 96.76% of female and 98.47% of male MOOC learners from Egypt chose to take at least one STEM MOOC.

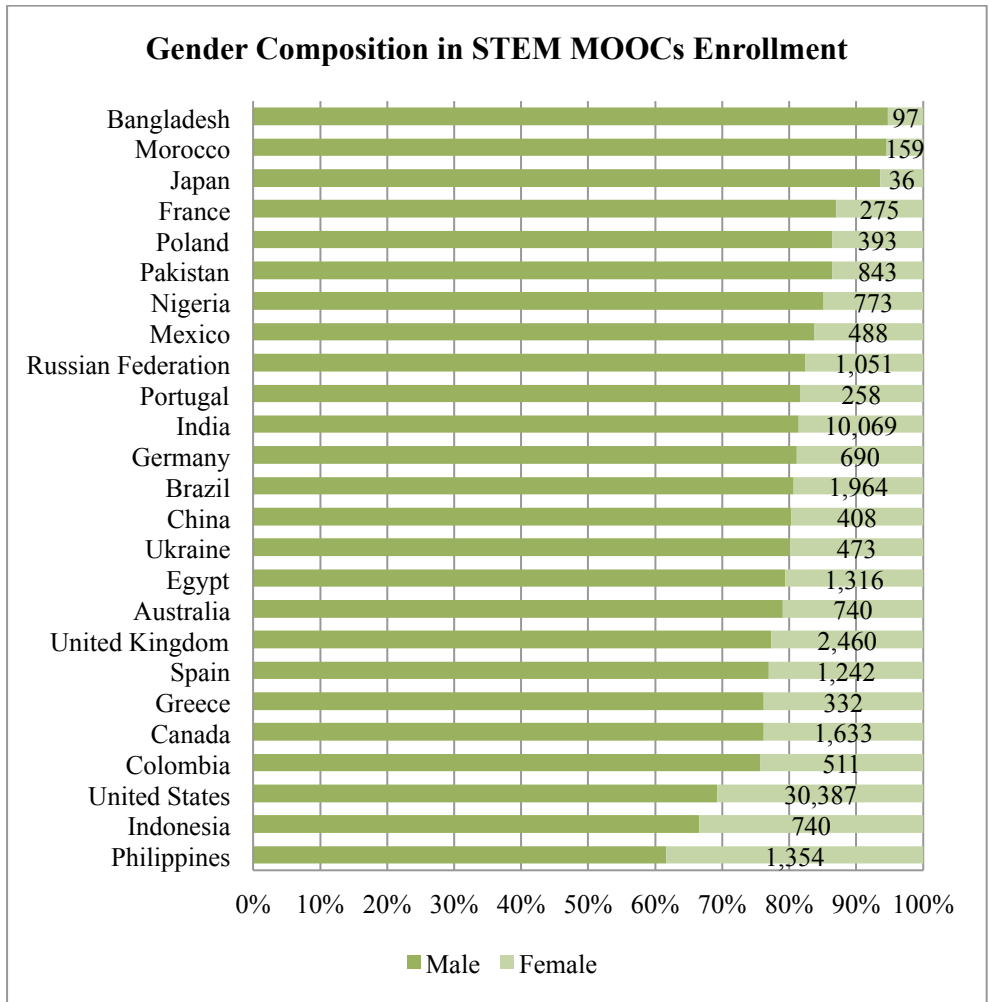

**Fig.1.** Gender Compositions in STEM MOOC Enrollment

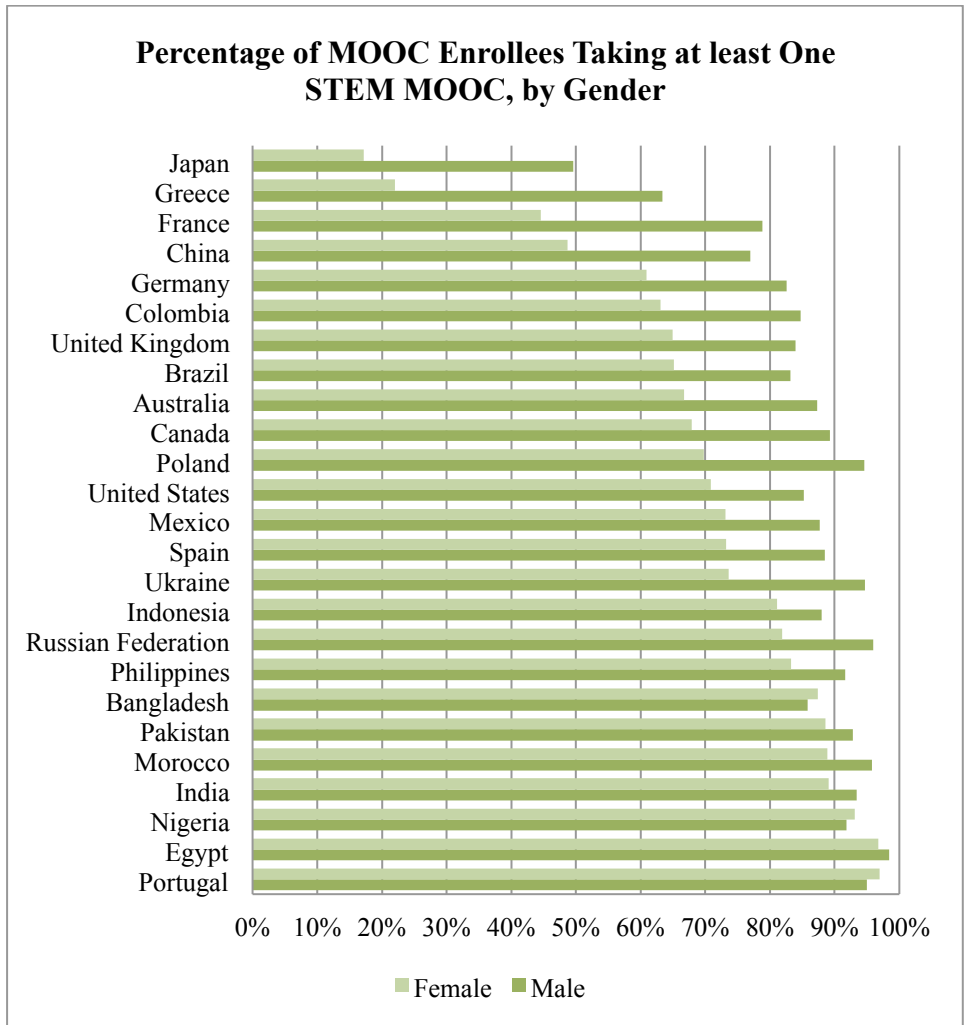

**Fig. 2.** Percentage of MOOC Enrollees Taking at least One STEM MOOC, by Gender

When it comes to the completion of STEM MOOCs (Figure 3), on average, only 2.98% of females who enrolled in STEM MOOCs actually completed at least one STEM MOOC compared with 3.01% of males. Furthermore, on average, 23.69% of STEM MOOC learners who completed a MOOC were female, but this varied greatly by country. Notably, females consisted of 52.7%, 46.43%, and 30.51% of STEM MOOC learners who completed at least one STEM MOOC in Indonesia, China, and the Philippines, respectively, as shown in Figure 4.

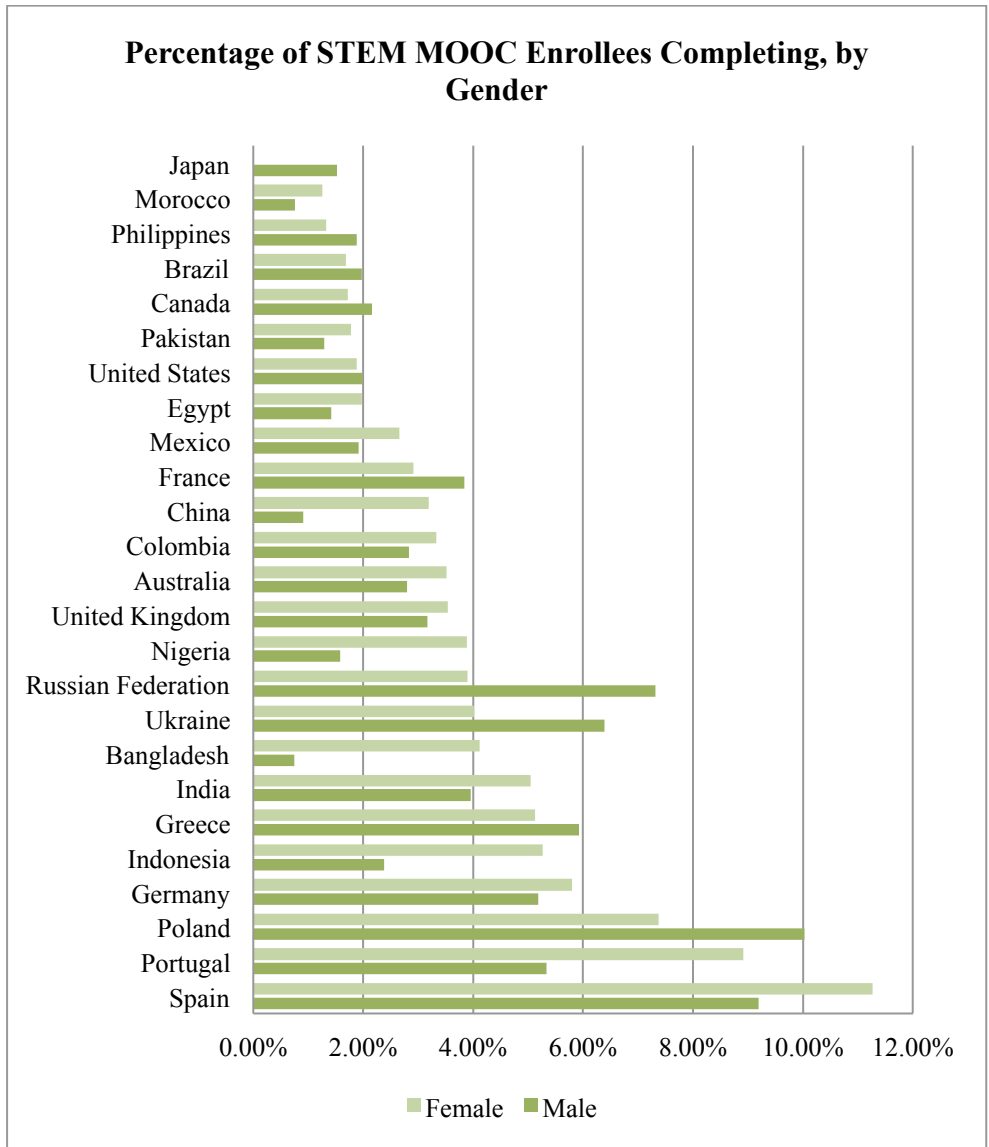

**Fig. 3.** Percentage of STEM MOOC Enrollees Completing, by Gender

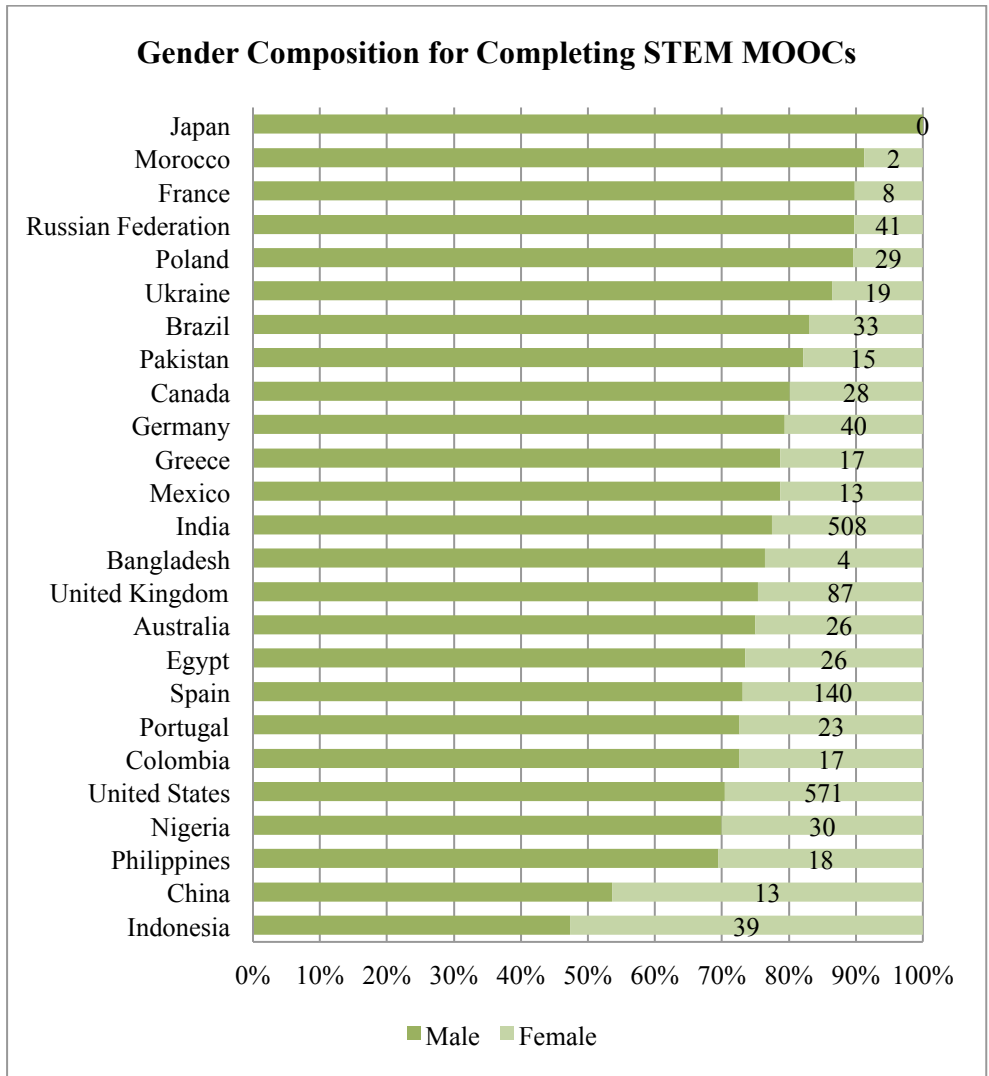

**Fig. 4.** Gender Composition for Completing STEM MOOCs

Across the whole sample, the odds of a female enrolling in at least one STEM MOOC was 62.7% less than that of a male (odds ratio = 0. 373, p < 0.001, when controlling for age and highest level of education of the individual), as shown by Model 2 in Table 1. Age (odds ratio = 0.978, p < 0.001) and highest level of education (odds ratio = 0.843, p < 0.001) were negatively associated with learners' enrollment in STEM MOOCs. A $10,000 increase of GDP per capita was associated with 15.5% decrease of the odds to take STEM MOOCs, when controlling for individual-level variables, as shown by Model 4 of Table 1. However, increased GDP per capita was associated with decreased enrollment (odds ratio = 0.846, p = 0.025) and increased gender gap (odds ratio = 0.908, p = 0.08), suggesting there is a higher rate and a smaller gender gap in STEM MOOC enrollment in less economically developed countries. In contrast, GGI was not significantly associated with the enrollment in a STEM MOOC (Model 3 in Table 1).

**Table 1**
Multilevel Logistic Regression on Whole Sample for Taking STEM MOOCs

|  | Model 1 Take STEM | Model 2 Take STEM | Model 3 Take STEM | Model 4 Take STEM | Model 5 Take STEM |
|---|---|---|---|---|---|
| Female | 0.396*** | 0.373*** | 0.373*** | 0.373*** | 0.453*** |
|  | (0.004) | (0.044) | (0.044) | (0.044) | (0.072) |
| Age |  | 0.978*** | 0.978*** | 0.978*** | 0.978*** |
|  |  | (0.001) | (0.001) | (0.001) | (0.001) |
| Education |  | 0.843*** | 0.843*** | 0.843*** | 0.843*** |
|  |  | (0.006) | (0.006) | (0.006) | (0.006) |
| GGI |  |  | 0.963 |  |  |
|  |  |  | (0.028) |  |  |
| GDP |  |  |  | 0.845* | 0.846* |
|  |  |  |  | (0.063) | (0.063) |
| Female*GDP |  |  |  |  | 0.908+ |
|  |  |  |  |  | (0.050) |
| N | 293144 | 280665 | 280665 | 280665 | 280665 |

Exponentiated coefficients; Standard errors in parentheses. The variable education is on the scale of 1-5; The variable GDP per capita is on the scale of 0-7 by dividing the raw GDP per capita by 10,000; The variable GGI is centered on the lowest value of GGI and is on the scale of 1-100 by multiplying the centered GGI by 100. The coefficients are odds ratio, and odds ratio of 1 means equivalent odds, above 1 means higher odds, and below 1 means lower odds.

+ p<0.1 * p<0.05 ** p<0.01 *** p<0.001

With regard to STEM MOOC completion, females and males were equally likely to complete STEM MOOCs (odds ratio = 1.093, p > 0.10), after controlling for age and highest level of education of the individual. However, when GGI and the interaction term between female and GGI were included in the model, females were more likely than males to complete STEM MOOCs (odds ratio = 1.765, p = 0.019), as shown by Model 5 in Table 2. Increased gender equality (GGI) was positively associated with the completion of STEM MOOCs (odds ratio = 1.066, p = 0.007). The interaction term between gender and GGI was negatively associated with completion of STEM MOOCs (odds ratio = 0.967, p = 0.034), indicating that females from the country with the lowest GGI (i.e. Pakistan) outperform males in STEM MOOCs completion, and this advantage decreases as GGI goes up. GDP was not statistically associated with learners' completion of STEM MOOCs. The findings suggest that females have the advantage in STEM MOOC completion in less gender-egalitarian countries.

**Table 2**
Multilevel Logistic Regression on Whole Sample for Completing STEM MOOCs

|  | Model 1 Complete STEM | Model 2 Complete STEM | Model 3 Complete STEM | Model 4 Complete STEM | Model 5 Complete STEM |
|---|---|---|---|---|---|
| Female | 1.127 | 1.093 | 1.093 | 1.092 | 1.765* |
|  | (0.101) | (0.100) | (0.100) | (0.100) | (0.428) |
| Age |  | 0.988*** | 0.988*** | 0.988*** | 0.988*** |
|  |  | (0.002) | (0.002) | (0.002) | (0.002) |
| Education | | 1.176*** | 1.177*** | 1.177*** | 1.177*** |
| | | (0.020) | (0.020) | (0.020) | (0.020) |
| GDP | | 1.079 | | | |
| | | (0.076) | | | |
| GGI | | | | 1.064** | 1.066** |
| | | | | (0.025) | (0.025) |
| Female*GGI | | | | | 0.967* |
| | | | | | (0.015) |
| N | 245700 | 234855 | 234855 | 234855 | 234855 |

Exponentiated coefficients; Standard errors in parentheses. The variable education is on the scale of 1-5; The variable GDP per capita is on the scale of 0-7 by dividing the raw GDP per capita by 10,000; The variable GGI is centered on the lowest value of GGI and is on the scale of 1-100 by multiplying the centered GGI by 100. The coefficients are odds ratio, and odds ratio of 1 means equivalent odds, above 1 means higher odds, and below 1 means lower odds.

+ p<0.1 * p<0.05 ** p<0.01 *** p<0.001

Different completion patterns were found between the Next Eleven counties and non Next Eleven countries. Female enrollees from the Next Eleven countries were much more likely to complete STEM MOOCs than males are (odds ratio =1.686, p = 0.004), after controlling for their age and highest level of education of the individual, as shown by Model 2 of Table 3. The increase of GDP (odds ratio = 1.754, p = 0.045) was positively associated with learners' completion of STEM MOOCs, as shown by Model 6 of Table 3. The interaction term between female and GDP was not statistically significant. GGI was not associated with the completion of STEM MOOCs. Further, when only looking at individuals in non Next Eleven countries, there is was no statistically significant relation between gender and STEM MOOC completion (Table 4).

**Table 3**
Multilevel Logistic Regression on Next Eleven Countries for Completing STEM MOOCs

| | Model 1 Completing STEM | Model 2 Completing STEM | Model 3 Completing STEM | Model 4 Completing STEM | Model 5 Completing STEM | Model 6 Completing STEM |
|---|---|---|---|---|---|---|
| Female | 1.649** | 1.686** | 1.683** | 2.211** | 1.707** | 1.906* |
| | (0.300) | (0.305) | (0.317) | (0.664) | (0.321) | (0.606) |
| Age | | 1.006 | 1.006 | 1.006 | 1.008 | 1.008 |
| | | (0.012) | (0.012) | (0.012) | (0.012) | (0.012) |
| Education | | 0.934 | 0.935 | 0.935 | 0.918 | 0.918 |
| | | (0.084) | (0.084) | (0.084) | (0.084) | (0.084) |
| GGI | | | 1.012 | 1.015 | | |
| | | | (0.014) | (0.014) | | |
| Female*GGI | | | | 0.973 | | |
| | | | | (0.024) | | |
| GDP | | | | | 1.712+ | 1.754* |
| | | | | | (0.472) | (0.492) |
| Female*GDP | | | | | | 0.735 |
| | | | | | | (0.528) |

| N | 28382 | 27530 | 27530 | 27530 | 27530 | 27530 |

Exponentiated coefficients; Standard errors in parentheses. The variable education is on the scale of 1-5; The variable GDP per capita is on the scale of 0-7 by dividing the raw GDP per capita by 10,000; The variable GGI is centered on the lowest value of GGI and is on the scale of 1-100 by multiplying the centered GGI by 100. The coefficients are odds ratio, and odds ratio of 1 means equivalent odds, above 1 means higher odds, and below 1 means lower odds.

+ p<0.1 * p<0.05 ** p<0.01 *** p<0.001

**Table 4**
Multilevel Logistic Regression on non Next Eleven Countries for Completing STEM MOOCs

|  | Model 1 Complete STEM | Model 2 Complete STEM | Model 3 Complete STEM | Model 4 Complete STEM | Model 5 Complete STEM |
|---|---|---|---|---|---|
| Female | 0.998 | 0.959 | 0.959 | 0.959 | 0.993 |
|  | (0.085) | (0.084) | (0.084) | (0.083) | (0.415) |
| Age |  | 0.987*** | 0.987*** | 0.987*** | 0.987*** |
|  |  | (0.002) | (0.002) | (0.002) | (0.002) |
| Education |  | 1.186*** | 1.186*** | 1.187*** | 1.187*** |
|  |  | (0.020) | (0.020) | (0.020) | (0.020) |
| GDP |  |  | 0.977 |  |  |
|  |  |  | (0.081) |  |  |
| GGI |  |  |  | 1.076* | 1.076* |
|  |  |  |  | (0.040) | (0.040) |
| Female*GGI |  |  |  |  | 0.998 |
|  |  |  |  |  | (0.026) |
| N | 217318 | 207325 | 207325 | 207325 | 207325 |

Exponentiated coefficients; Standard errors in parentheses. The variable education is on the scale of 1-5; The variable GDP per capita is on the scale of 0-7 by dividing the raw GDP per capita by 10,000; The variable GGI is centered on the lowest value of GGI and is on the scale of 1-100 by multiplying the centered GGI by 100. The coefficients are odds ratio, and odds ratio of 1 means equivalent odds, above 1 means higher odds, and below 1 means lower odds.

+ p<0.1 * p<0.05 ** p<0.01 *** p<0.001

**Discussion and conclusions**

Complement to previous work investigating the democratization of MOOCs in the US (Hansen & Reich, 2015), we found that globally MOOCs have the potential to provide learning opportunities for females in less developed countries. Findings from this study support the hypothesis that greater gender segregation may exist in more economically developed countries. Though it is unclear from the current findings why this is, evidence from other studies point to cultural beliefs (Charles & Bradley, 2009). Free and easy access of MOOCs in developing countries allow females to try out STEM courses that are not easily available to them in their local communities. The online courses may provide opportunities that are not open to females in their countries. Females from developing countries may be motivated by the financial rewards to complete STEM MOOCs (Khazan, 2014). However, we suggest future studies be conducted to understand female's decisions to enroll in and complete STEM MOOCs.

The study suggests that MOOCs may be providing opportunities for females to take STEM courses, especially females from less gender-egalitarian and less economically developed countries. MOOCs might bring broad country-level social benefits for less socially and economically developed countries. Considering that females in our sample consisted of only 23.89 % of STEM MOOC learners, we suggest that more outreach should be taken to promote MOOC participation among females. Females may be less aware of the options of taking STEM courses online for free and the opportunities or financial rewards that could result from taking these courses (Eccles, 1994; Chen et al., 2015).

**Acknowledgement**

We are very grateful for the public MOOC dataset provided by HarvardX-MITx.